\documentclass{ws-procs961x669}            
\usepackage{color}
\newcommand{\eq}[1]{(\ref{#1})}
\begin{document}
\title{Cosmic Inflation from Emergent Spacetime Picture}

\author{Hyun Seok Yang}

\address{Center for Quantum Spacetime, Sogang University, Seoul
121-741, Korea\\
$^*$E-mail: hsyang@sogang.ac.kr}

\begin{abstract}
We argue that the emergent spacetime picture admits a background-independent formulation
of cosmic inflation. The inflation in this picture corresponds to the dynamical emergence of
spacetime while the conventional inflation is simply an (exponential) expansion of
a preexisting spacetime owing to the vacuum energy carried by an inflaton field.
We show that the cosmic inflation arises as a time-dependent solution of the matrix
quantum mechanics describing the dynamical process of Planck energy condensate in vacuum
without introducing any inflaton field as well as an {\it ad hoc} inflation potential.
Thus the emergent spacetime picture realizes a background-independent description of
the inflationary universe which has a sufficiently elegant and explanatory power to defend
the integrity of physics against the multiverse hypothesis.
\end{abstract}


\bodymatter

\section{Why Is Emergent Spacetime Necessary for Cosmic Inflation?}

Like black holes, the big bang in the very early universe involves extreme conditions that
neither relativity nor quantum theory can explain on its own. If we trace back the history
of our universe, we will meet an initial singularity in which matter reached almost infinite density
and then, according to the theory of general relativity, the space was contracted to a point
and the time flow nearly stopped. Thus the big bang suffers the initial singularity
in which space and time cease to exist. This implies that the big bang must be
a cosmological event generating space and time as well as matters. The initial singularity
cannot be avoided in the inflationary cosmology either because there has to be
a definite beginning to an inflationary universe.\cite{bgv} This means that the inflation
is incomplete to describe the very beginning of our universe and some new physics is
needed to probe the past boundary of the inflating regions.
One possibility is that there must have been some sort of quantum creation event
as a beginning of the universe.

The inflation scenario so far has been formulated in the context of effective field theory
coupled to general relativity. Hence, in this scenario, the existence of space
and time is {\it a priori} assumed from the beginning and the scenario only describes
what happens in a given spacetime. In other words, the inflationary scenario does not describe
any generation (or creation) of spacetime but simply characterizes an expansion of
a preexisting spacetime. It never addresses the (dynamical) origin of spacetime.
Moreover, in most inflationary models, once inflation happens, it never stops and produces
not just one universe, but an infinite number of universes. Therefore the conventional inflation
inevitably leads to the multiverse which is not falsifiable since the multiverse
cannot be tested experimentally.\cite{multiverse}

In consequence, we need a consistent quantum theory to describe how spacetime was generated
through the big bang in order to correctly understand the origin of spacetime and our universe.
In particular, we need a background-independent theory which does not assume the prior existence
of spacetime background but instead provides a mechanism of spacetime generation such that
any spacetime structure including flat spacetime arises as a solution of the theory itself.
In other words, we need a background-independent formulation of cosmic inflation
in which the inflation corresponds to the dynamical emergence of spacetime.
Recently such a background independent formulation was proposed \cite{part1} so that
the cosmic inflation arises as a time-dependent solution of the matrix
quantum mechanics describing the dynamical process of Planck energy condensate in vacuum
without introducing any inflaton field as well as an {\it ad hoc} inflation potential.
The underlying mathematical principle is the well-known duality between geometry and algebra.
In this scheme, the noncommutative (NC) algebra plays a more fundamental role from which
the spacetime geometry is derived and the emergent gravity from NC $U(1)$ gauge theory
is basically the large $N$ duality, as depicted in Fig. \ref{fchart:lnd}:

\begin{figure}
\begin{center}
\includegraphics[width=4.5in]{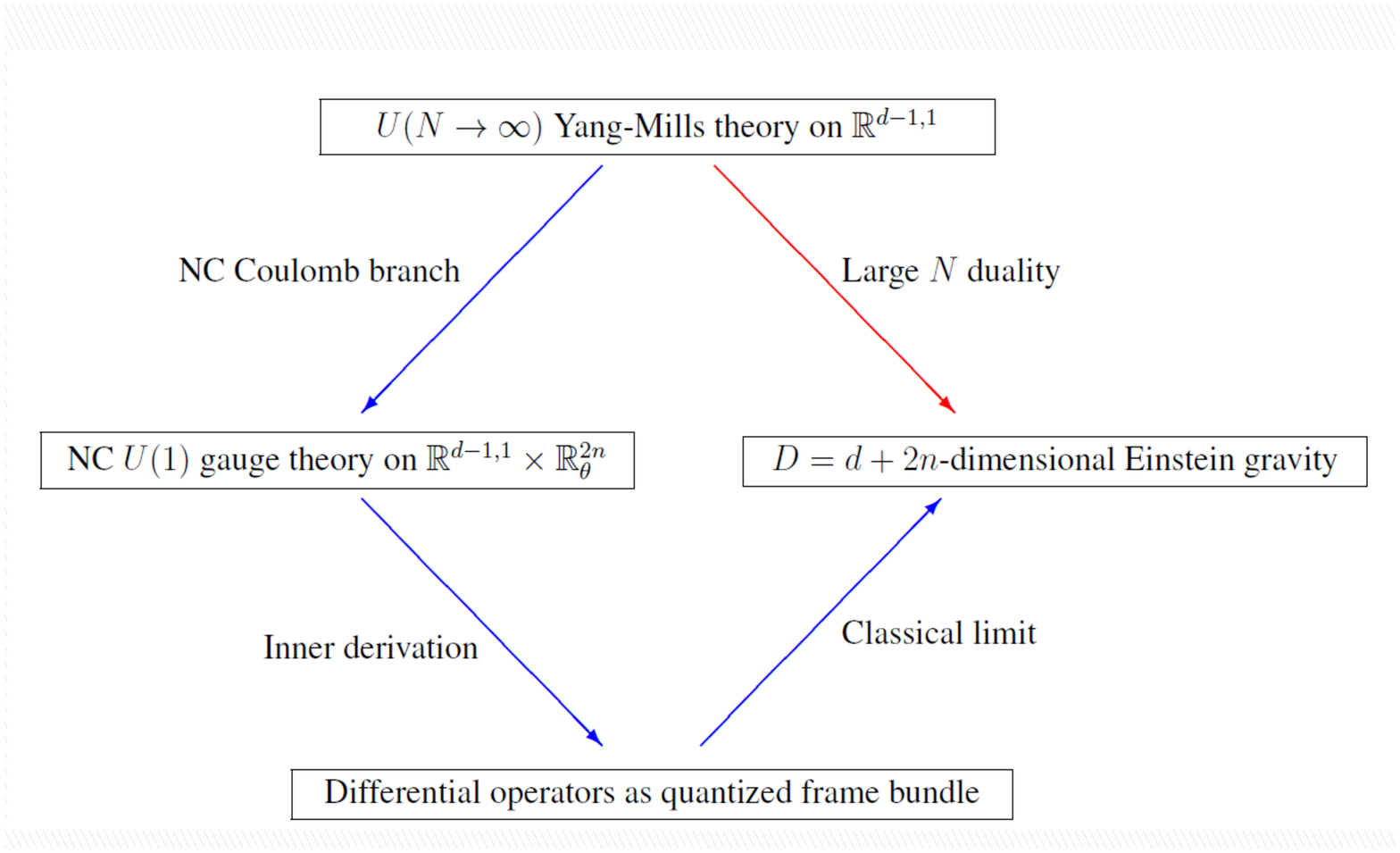}
\end{center}
\caption{Flowchart for large $N$ duality.\cite{part1}}
\label{fchart:lnd}
\end{figure}

Since the concept of the multiverse raises deep conceptual issues even to require to change our
view of science itself, it should be important to ponder on the real status of the multiverse
whether it is simply a mirage developed from an incomplete physics like the ether
in the late 19th century or it is of vital importance even in more complete theories.
In next section, we will illuminate \cite{part1} how the emergent spacetime picture brings
about radical changes of physics, especially, regarding to physical cosmology.

\section{Inflationary Universe from Emergent Spacetime}

Let us consider the matrix quantum mechanics (MQM) to address the background-independent
formulation of cosmic inflation. The action in this case is given by \cite{wtaylor}
\begin{eqnarray} \label{mqm}
 S &=& \frac{1}{g^2} \int dt \, \mathrm{Tr} \Bigl( \frac{1}{2} (D_0 \phi_a)^2
 + \frac{1}{4}[\phi_a, \phi_b]^2 \Bigr) \nonumber \\
   &=& \frac{1}{4 g^2} \int dt \, \eta^{AC} \eta^{BD} \mathrm{Tr}
   [\phi_A, \phi_B] [\phi_C, \phi_D],
\end{eqnarray}
where $\phi_0 \equiv i D_0 = i \frac{\partial}{\partial t} + A_0 (t), \; \phi_A (t) = (\phi_0, \phi_a) (t)$
and $\eta^{AB} = \mathrm{diag} (-1, 1, \cdots, 1)$, $A, B = 0, 1, \cdots, 2n.$
With the notation of the symbol $\eta^{AB}$, it is easy to see that the matrix action \eq{mqm}
has a global automorphism given by
\begin{equation}\label{poincare-auto}
    \phi_A \to \phi'_A = {\Lambda_A}^B \phi_B + c_A
\end{equation}
if ${\Lambda_A}^B$ is a rotation in $SO(2n,1)$ and $c_A$ are constants proportional
to the identity matrix. It turns out \cite{part1} that the global symmetry \eq{poincare-auto}
is responsible for the Poincar\'e symmetry of flat spacetime emergent from a vacuum
in the Coulomb branch of MQM and so will be called the Poincar\'e automorphism.

We remark that the time $t$ in the action \eq{mqm} is not a dynamical variable but a parameter.
The concept of time has been introduced in Ref. \refcite{part1}
by considering a one-parameter family of deformations
of zero-dimensional matrices which is parameterized by the coordinate $t$.
Then the one-parameter family of deformations can be regarded as the time-evolution of
a dynamical system. A close analogy with quantum mechanics implies that
the concept of emergent time is connected with the time-evolution of the dynamical system.
In this context, the one-dimensional matrix model \eq{mqm} can be interpreted as
a Hamiltonian system of a zero-dimensional (e.g., IKKT) matrix model.\cite{q-emg}

The equations of motion for the matrix action \eq{mqm} are given by
\begin{equation}\label{eom-mqm}
    D_0^2 \phi_a + [\phi_b, [\phi_a, \phi_b]] = 0,
\end{equation}
which must be supplemented with the Gauss constraint
\begin{equation}\label{gauss-mqm}
    [\phi_a, D_0 \phi_a] = 0.
\end{equation}
Now we want to apply the large $N$ duality to MQM that is the $d=1$ case in Fig. \ref{fchart:lnd}.
It is important to notice that there are two different kinds of vacua in the Coulomb branch
if we consider the $N \to \infty$ limit. In addition to the conventional commutative vacuum
obeying the property $[\phi_a, \phi_b]|_{\mathrm{vac}} = 0$, there exists a novel coherent vacuum,
the so-called NC Coulomb branch \cite{part1}, defined by
\begin{equation}\label{nc-coulomb}
    [\phi_a, \phi_b]|_{\mathrm{vac}} = - i B_{ab} \qquad \Rightarrow \qquad
    \langle \phi_a \rangle_{\mathrm{vac}} = p_a \equiv B_{ab} y^b
\end{equation}
where the vacuum moduli $y^a$ satisfy the Moyal-Heisenberg algebra given by
\begin{equation}\label{extra-nc2n}
    [y^a, y^b] = i \theta^{ab}, \qquad a, b = 1, \cdots, 2n
\end{equation}
and $(\theta)^{ab} = (B^{-1})^{ab} = \alpha' (\mathbf{1}_n \otimes i \sigma^2)$ is a $2n \times 2n$
constant symplectic matrix. We emphasize that the NC Coulomb branch \eq{nc-coulomb}
together with a constant vacuum energy density given
by $\mathcal{E} \equiv \langle A_0 (t) \rangle_{\mathrm{vac}}$ is a consistent {\it vacuum} solution
of MQM since it satisfies the equations of motion $\eq{eom-mqm}$
as well as the Gauss constraint \eq{gauss-mqm}.
Since $\mathcal{E}$ is proportional to the identity matrix, it plays no role in the temporal
covariant derivative $D_0$ and so it can be dropped without loss of generality.
Also note that the coherent vacuum \eq{nc-coulomb} saves the NC nature
of matrices while the conventional vacuum dismisses the property.

If we consider fluctuations around the NC vacuum \eq{nc-coulomb} given by
\begin{equation}\label{gen-sol}
 D_0 = \frac{\partial}{\partial t} - i \widehat{A}_0 (t, y),
 \qquad \phi_a = p_a + \widehat{A}_a (t, y),
\end{equation}
the action for the fluctuations is exactly mapped to a $(2n+1)$-dimensional NC $U(1)$ gauge
theory as indicated in Fig. \ref{fchart:lnd}. If the conventional commutative vacuum were chosen,
we would have failed to realize the higher-dimensional NC $U(1)$ gauge theory from MQM.
Indeed it turns out \cite{part1,q-emg} that the NC Coulomb branch is crucial to realize
the emergent gravity from matrix models or large $N$ gauge theories as summarized
in Fig. \ref{fchart:lnd}. Furthermore the NC Coulomb branch \eq{nc-coulomb} is very different
from the conventional commutative vacuum because the former carries a nonzero vacuum energy
density in contrast to the latter for which the vacuum energy density identically vanishes.
Using the higher-dimensional NC $U(1)$ gauge theory, we can calculate
the energy density for the vacuum condensate in the NC Coulomb branch defined by
\begin{equation}\label{nc-vacuum}
    \langle \phi_A \rangle_{\mathrm{vac}} = p_A
    = \Big( i \frac{\partial}{\partial t}, p_a \Big).
\end{equation}
The result is given by
\begin{equation}\label{vacuum-density}
    \rho_{\mathrm{vac}} = \frac{1}{4G_{YM}^2} |B_{ab}|^2
\end{equation}
where $G_{YM}^2 = (2\pi)^n |\mathrm{Pf} \theta| g^2$ is the $(2n+1)$-dimensional gauge
coupling constant.

It was shown in Ref. \refcite{part1} that quantum gravity can be derived from
the electromagnetism on NC spacetime by realizing the large $N$ duality in Fig. \ref{fchart:lnd}
via the duality chain given by
\begin{equation}\label{dd-chain}
  \mathcal{A}^d_N  \quad \Longrightarrow \quad \mathcal{A}^d_\theta \quad
  \Longrightarrow \quad \mathfrak{D}^d.
\end{equation}
For the $d=1$ case in Fig. \ref{fchart:lnd}, the dynamical variables in MQM take values
in $\mathcal{A}_N^1$ while those in $D=(2n+1)$-dimensional NC $U(1)$ gauge theory
take values in $\mathcal{A}_\theta^1$. These two NC algebras $\mathcal{A}_N^1$ and
$\mathcal{A}_\theta^1$ are related to each other by considering
the NC Coulomb branch \eq{nc-vacuum}.\cite{part1}
The module of derivations $\mathfrak{D}^1$ is a direct sum of
the submodules of horizontal and inner derivations:\cite{azam}
\begin{equation}\label{d-deriv}
  \mathfrak{D}^1 = \mathrm{Hor}(\mathcal{A}^1_N) \oplus \mathfrak{D} (\mathcal{A}^1_N) \cong
  \mathrm{Hor}(\mathcal{A}^1_\theta) \oplus \mathfrak{D} (\mathcal{A}^1_\theta),
\end{equation}
where horizontal derivation is locally generated by a vector field
\begin{equation}\label{dhor-vec}
k (t, y) \frac{\partial}{\partial t} \in \mathrm{Hor}(\mathcal{A}^1_\theta).
\end{equation}
In particular we are interested in the derivation algebra generated by the dynamical variables
in Eq. \eq{gen-sol}. It is defined by
\begin{equation}\label{der-d}
    \widehat{V}_A = \{ \mathrm{ad}_{\phi_A} = -i [\phi_A, \; \cdot \;] | \phi_A (t,y)
    = (i D_0, \phi_a) (t,y) \}    \in \mathfrak{D}^1.
\end{equation}

The large $N$ duality in Fig. \ref{fchart:lnd} says that the gravitational variables such as
vielbeins in general relativity arise from the commutative limit of NC $U(1)$ gauge fields
via the map \eq{der-d}. Then one may ask where flat Minkowski spacetime comes from.
It turns out \cite{hsy-jhep09,hsy-review} that the $(2n+1)$-dimensional flat Minkowski spacetime
is emergent from the vacuum condensate \eq{nc-vacuum} since the corresponding vielbeins
and the metric are given by $E^{(0)}_A =  \widehat{V}^{(0)}_A = \Big( \frac{\partial}{\partial t},
\frac{\partial}{\partial y^a} \Big)$ and $ds^2 = -dt^2 + d \mathbf{y} \cdot d \mathbf{y}$.
A striking fact is that the flat Minkowski spacetime is emergent from the Moyal-Heisenberg
algebra \eq{nc-coulomb} whose energy density is given by Eq. \eq{vacuum-density}.
Thus the flat spacetime was originated from the uniform vacuum energy \eq{vacuum-density}
known as the cosmological constant in general relativity.
This is a tangible difference from Einstein gravity, in which
the energy-momentum tensor identically vanishes, i.e. $T_{AB} = 0$, for the flat spacetime.
However, since we have started with the matrix model \eq{mqm} in which any spacetime structure
has not been assumed in advance, the spacetime was not existent at the beginning
but simply emergent from the vacuum condensate \eq{nc-coulomb}. Therefore the
Planck energy condensation into vacuum must be regarded as a dynamical process.

It is not difficult to show \cite{part1} that the dynamical process for the vacuum condensate
is described by the time-dependent vacuum configuration given by
\begin{equation}\label{time-vacuum}
    \langle \phi_a (t) \rangle_{\mathrm{vac}} = p_a (t) = e^{\frac{\kappa t}{2}} p_a, \qquad
    \langle \widehat{A}_0 (t, y) \rangle_{\mathrm{vac}} = \widehat{a}_0 (t, y),
\end{equation}
where the temporal gauge field is given by an open Wilson line
\begin{equation}\label{open-wilson}
 \widehat{a}_0 (t, y) = \frac{\kappa}{2} \int_0^1 d \sigma \frac{d y^a (\sigma)}{d \sigma} p_a (\sigma)
\end{equation}
along a path parameterized by the curve $y^a (\sigma) = y_0^a + \zeta^a (\sigma)$
where $\zeta^a (\sigma) = \theta^{ab} k_b \sigma$ with $0 \leq \sigma \leq 1$
and $y^a (\sigma=0) \equiv y_0^a$ and $y^a (\sigma=1) \equiv y^a$.
The constant $H \equiv (n-1) \kappa$ is identified with the inflationary Hubble constant.
The $(2n+1)$-dimensional basis for the time-dependent vacuum \eq{time-vacuum} can easily
be calculated using the map \eq{der-d}:
\begin{equation}\label{vac-vielbein}
    V_0 = \frac{\partial}{\partial t} - \frac{\kappa}{2} y^a \frac{\partial}{\partial y^a},
    \qquad V_a = e^{\frac{\kappa t}{2}} \frac{\partial}{\partial y^a}.
\end{equation}
And the dual orthonormal one-forms are given by
\begin{equation}\label{vac-dual}
    e^0 = dt, \qquad e^a = e^{Ht} dy_t^a
\end{equation}
where $y_t^a = e^{\frac{\kappa t}{2}} y^a$.
Here we used the relation $V_A = (E_0, \lambda E_a)$,
and $\lambda = e^{ n \kappa t}$ for the vacuum configuration \eq{time-vacuum}.
In the end, the time-dependent metric
for the inflating background \eq{time-vacuum} is given by \cite{part1}
\begin{equation}\label{inflation}
    ds^2 = - dt^2 + e^{2Ht} d\mathbf{y}_t \cdot d\mathbf{y}_t.
\end{equation}
Note that the temporal gauge field \eq{open-wilson} is crucial to satisfy Eqs. \eq{eom-mqm}
and \eq{gauss-mqm} and to get a geodesically complete inflationary spacetime.
Moreover the metric \eq{inflation} is conformally flat, i.e., the corresponding Weyl tensors
identically vanish and so describes a homogeneous and isotropic inflationary universe
known as the Friedmann-Robertson-Walker metric in physical cosmology.

It is also easy to get a general Lorentzian metric describing $(2n + 1)$-dimensional inflating
spacetime by considering arbitrary fluctuations around the inflationary
background \eq{time-vacuum}.\cite{part1}
They form a time-dependent NC algebra given by
\begin{equation}\label{time-ncalg}
    {}^t\mathcal{A}_\theta^1 \equiv \Big\{ \widehat{\phi}_0 (t, y) = i \frac{\partial}{\partial t}
    + \widehat{A}_0 (t, y), \quad \widehat{\phi}_a (t, y) = e^{\frac{\kappa t}{2}} \big( p_a
    + \widehat{A}_a (t, y) \big) \Big\}.
\end{equation}
We denote the corresponding time-dependent matrix algebra by ${}^t\mathcal{A}_N^1$ which
consists of a time-dependent solution of the action \eq{mqm}.
Then the general Lorentzian metric describing a $(2n + 1)$-dimensional inflationary universe
can be obtained by the following duality chain:
\begin{equation}\label{inf-chain}
  {}^t\mathcal{A}^1_N  \quad \Longrightarrow \quad {}^t\mathcal{A}^1_\theta \quad
  \Longrightarrow \quad {}^t\mathfrak{D}^1.
\end{equation}
The module ${}^t\mathfrak{D}^1$ of derivations of the NC algebra ${}^t\mathcal{A}_\theta^1$ is
given by
\begin{equation}\label{inf-der}
    {}^t\mathfrak{D}^1 = \Big\{ \widehat{V}_A (t) = (\widehat{V}_0, \widehat{V}_a) (t) |
    \widehat{V}_0 (t) = \frac{\partial}{\partial t} + \mathrm{ad}_{\widehat{A}_0},
    \quad \widehat{V}_a (t) = e^{\frac{\kappa t}{2}} \mathrm{ad}_{\widehat{\phi}_a} \Big\},
\end{equation}
where the adjoint operations are defined by Eq. \eq{der-d}. In the classical limit of
the module \eq{inf-der}, we get a general inflationary universe described by
\begin{equation}\label{gen-inf}
    ds^2 = -dt^2 + \lambda^2  e^{2Ht} v^a_\alpha v^a_\beta (dy_t^\alpha - \mathbf{A}^\alpha)
    (dy_t^\beta - \mathbf{A}^\beta)
\end{equation}
where the conformal factor $\lambda = 1 + \delta\lambda$ is determined by the volume-preserving condition.

In conclusion, the cosmic inflation arises as a time-dependent solution of
a background-independent theory describing the dynamical process of Planck energy condensate
in vacuum without introducing any inflaton field as well as an {\it ad hoc} inflation potential.
The large $N$ duality in Fig. \ref{fchart:lnd} also implies \cite{part1} that cosmic inflation
triggered by the Planck energy condensate into vacuum must be a single event.
Thus the emergent spacetime is a completely new paradigm so that the multiverse debate
in physics circles has to seriously take it into account.\cite{debate}




\end{document}